\documentclass[fleqn,usenatbib]{mnras}

\usepackage{newtxtext,newtxmath}

\usepackage[T1]{fontenc}

\DeclareRobustCommand{\VAN}[3]{#2}
\let\VANthebibliography\thebibliography
\def\thebibliography{\DeclareRobustCommand{\VAN}[3]{##3}\VANthebibliography}

\usepackage{graphicx}	
\usepackage{amsmath}	

\begin{document}
\title[$\gamma^{2}$ Vel Speckle]{Speckle Imaging of $\gamma^{2}$ Velorum: The Inner Wind Possibly Resolved}

\author[M. Shara et al.]{Michael M. Shara$^{1}$\thanks{E-mail: mshara@amnh.org}, Steve B. Howell$^{2}$, Elise Furlan$^{3}$,  James T. Garland$^{1}$, Anthony F.J. Moffat$^{4}$, 
\newauthor {and David Zurek$^{1}$}
\\
$^{1}$Department of Astrophysics, American Museum of Natural History, Central Park West at 79th Street, New York, NY 10024, USA\\
$^{2}$NASA Ames Research Center, Moffett Field, CA 94035, USA\\
$^{3}$NASA Exoplanet Science Institute, Caltech/IPAC, Mail Code 100-22, 1200 E. California Blvd., Pasadena, CA 91125, USA\\
$^{4}$D\'epartement de Physique et Centre de Recherche en Astrophysique du Qu\'ebec, Universit\'e de Montr\'eal, Montr\'eal, QC H3C 3J7, Canada
}

\date{Accepted XXX. Received YYY; in original form ZZZ}

\pubyear{2023}


\label{firstpage}
\pagerange{\pageref{firstpage}--\pageref{lastpage}}
\maketitle

\begin{abstract}
Accurately quantifying the rates dM/dt at which massive stars lose mass is essential to any understanding of their evolution. All dM/dt estimates to date assume wind clumping factors; not allowing for clumping leads to overestimates of dM/dt and underestimates of lifetimes and masses when these stars explode as supernovae. Mid-IR spectroscopy suggested that the wind of the nearest Wolf-Rayet star, $\gamma^{2}$ Vel, is resolved with a Full Width at 10 per cent intensity of 0.5 arcsec, or 171 AU at the 342 pc distance of the star. As the Zorro speckle imager on Gemini-South is capable of $\sim$ 0.02 arcsec resolution, we have used it to image $\gamma^{2}$ Vel at two orbital phases (0.30 and 0.44) with two narrowband and two intermediate-band filters in an attempt to resolve its wind. Our observations demonstrate that $\gamma^{2}$ Vel's wind may be resolved as a $\sim$ 0.07 arcsec westward elongation through an 832 nm filter at orbital phase 0.3. If confirmed, this is the smallest scale ($\sim$24 AU) at which a WR star wind asymmetry has been directly imaged. Similar imaging at multiple phases is needed to determine if the asymmetry is due to stochastic wind clumping, co-rotating interaction regions or colliding-wind, cone-shaped shocks.
\end{abstract}



\section{Introduction and Motivation}

 Wolf-Rayet (WR) stars are the rapidly-evaporating descendants of O-type stars. Their large initial masses, short lifetimes ($\sim$300 kyr) due to enormous mass-loss rates, and concomitant strong emission lines make them easily detectable tracers of recent star formation through much of our Galaxy \citep{Shara2009,Kanarek2015}, the Local Group \citep{Moffat1983,Massey1987,Massey2017}, and even as distant as M101 \citep{Shara2013}. Nitrogen-rich (WN type) and Carbon or Oxygen-rich (WC and WO type) WR stars likely die as type Ib and Ic supernovae, respectively \citep{Gaskell1986,Corsi2012}. A significant fraction of type Ib and Ic SNe are believed to be produced by the wind-stripped cores of WR stars \citep{Dessart2020,Woosley2021}. WR stars thus play important roles in understanding the late stages of single and binary star evolution and supernovae \citep{Crowther2007}.

$\gamma^{2}$ Velorum is a WR-O star binary system with a period of 78.53 d, inclination of $\sim$ 65 degrees, and a projected angular semimajor axis of $\sim$3.57 mas \citep{North2007}. Very Large Telescope Interferometer/Astronomical Multi-BEam Recombiner (VLTI/AMBER) data yielded the spectral components of both stars \citep{Lamberts2017}. These observations confirmed the WR star's WC8 classification \citep{DeMarco1999}, showed that the O star spectrum is peculiar within its class, and that the $\sim 30\,$M$_\odot$ O star and its WC companion display an orbital separation of 0.8 - 1.6 AU and a wind-collision zone at about 1.2 AU.

At a \textit{Gaia}-determined distance of 342 pc, the WR component of the WC8 + O binary \citep{DeMarco1999} is about four times closer to Earth than the next closest WR star. The WR star has a mass of 9 M$_\odot$ which radiates $1.7\times10^{5}\,$L$_\odot$ \citep{North2007}. It is blowing off its ionized, mainly Helium + Carbon atmosphere at a rate of $\sim$ $1.5\times10^{-5}\,$M$_\odot$/yr, about 100X the ejection rate of its O-type companion \citep{deMarco2000}. There is no closer star in the sky that is consistently ejecting mass at anywhere near $\gamma^{2}$ Velorum’s rate, and the wind of the two stars is overwhelmingly dominated by its WR star. There is thus is no better target in the sky which offers the possibility of directly resolving the wind of a WR star on smaller physical scales than heretofore accomplished. 

In a pioneering study, \citet{Roche2012} determined, via narrow-band
mid-IR spectroscopy, that the $[\ion{S}{IV}]$ 10.52 micron and $[\ion{Ne}{II}]$ 12.81 micron emission region surrounding $\gamma^2$ Vel is spatially extended, with a claimed Full Width at 10 per cent Intensity of 0.5 arcsec (see their Fig. 2), compared to the nearby continuum which is unresolved. The observed emission-line distributions were azimuthally symmetric, suggestive of a spherically symmetric outflow. A significant degree of density clumping (a factor $\sim$ 10) in a model wind provided a better match to the spectrographic observations than an un-clumped model. These results of \citet{Roche2012} are further motivation to try to resolve the wind of $\gamma^{2}$ Vel. While its 20 mas resolution does not enable the Zorro speckle imager mounted at the Gemini-South telescope to resolve the O and WC8 star, or the inner wind shock cone, the spiral pattern predicted by their outflowing wind model extends far beyond the binary \citep{Lamberts2017}. 

 Zorro provides simultaneous red and blue-band, diffraction-limited optical imaging (FWHM $\sim$ 0.02 arcsec at H$\,\alpha$ and 0.015 arcsec at 466 nm). (See \citet{Scott2018,Scott2021} and the `Alopeke-Zorro Web pages{\footnote{https://www.gemini.edu/instrumentation/alopeke-zorro}} for full descriptions of the camera and its filters). Close binaries (with 0.02 - 0.1 arcsec separation) and brightness differences up to 5 magnitudes, and wider companions (0.1 to 1.2 arcsec) with 8-10 magnitude differences are routinely resolved by Zorro and its ``twin'' speckle camera 'Alopeke at the Gemini-North telescope \citep{Horch2019,Lester2021,Anguita-Aguero2022,Shara2022}. 

Features of order 0.5 arcsec in angular extent should be immediately visible via inspection of Gemini South's Zorro images, enabling a direct, model-independent determination of clumping and/or asymmetry in $\gamma^{2}$ Vel's inner wind – a significant “first” for WR stars. We thus requested and were granted 1.5 hours of Gemini-South/Zorro observing time to attempt direct resolution of $\gamma^{2}$ Vel's wind.

In section 2 we describe the data acquired and their reductions. In section 3 we present our results and the implications. We briefly summarize our results in section 4. 

\section{Observations and Data Reductions}

The four Zorro filters used in this study have central wavelengths/FWHM which are: 466/44 nm, 656.5/3.2 nm, 716/52 nm and 832/40 nm. The 466 nm filter encompasses the emission features of the WC8 WR star due to $\ion{He}{II}$ 468.6 nm and $\ion{C}{III}$ 464.0 nm \citep{Richardson2017}; the 656.5 nm filter is nominally a hydrogen H$\,\alpha$ filter, but it also captures the Pickering $\ion{He}{II}$ emission line at 656.0 nm; and the 716 nm and 832 nm intermediate-bandpass filters capture weaker $\ion{C}{II}$ and $\ion{C}{III}$ lines, respectively \citep{Vreux1983}. 

At V = 1.83 $\gamma^{2}$ Vel is the brightest WR star in the sky. On 2023 January 15 (orbital phase 0.30), $\gamma^{2}$ Vel was observed at Gemini-South with the Zorro speckle camera through the above four filters. Each frame was taken at the usual exposure time of 60 msec. The H$\,\alpha$ and 832 nm observations had a dozen or so frames that showed saturation - these were eliminated prior to the reductions. The 466 and 716 nm data were saturated and unusable. On 2023 January 26 (orbital phase 0.44) $\gamma^2$ Vel was re-observed in the 466 and 716 nm filters at 10 msec frame times, yielding fully usable data. The total exposure time in each of the four filters is 2400 s. 

Each set of one-minute-long standard star observations was applied as a separate PSF standard to the target. On January 15, three sets of observations of HD 67704 (2 sets in H$\,\alpha$ + 832 nm and 1 set in 466 nm + 716 nm) and 4 sets of HR 3179 (466 nm + 716 nm) were taken, while on January 26, 6 sets of HD 67704 observations (466 nm + 716 nm) and 6 sets of HR 3179 (466 nm + 716 nm) were taken.

Seeing was typically $\sim$ 0.6 arcsec during all observations, which were carried out at airmass < 1.1. Sky conditions were gray to dark, with the lunar phase just past last quarter on 2023 January 15, and 4 days past new moon on 2023 January 26.

The data reduction methodology and final data products are described in \citet{Horch2012} and \citet{Howell2011}, respectively. The data were reduced in several ways - using each of the two Point Spread Function standards as well as different subsets of the PSF frames. Both standards were used to separately reduce all of the data, with similar results in each case.
\section{RESULTS} 

Since we are looking for potentially subtle variations and asymmetries in the brightness distribution of $\gamma^{2}$ Vel, we must be certain that our calibrating stars are indeed pointlike, and that the reduction procedure does not introduce artifacts. We therefore observed {\it two} standard stars close on the sky to $\gamma^{2}$ Vel, and as noted above, reduced each one both as the target star and as the PSF standard.

\begin{figure*}
\vspace{5 cm}
\hspace{-1 cm}
\includegraphics[width=20 cm]{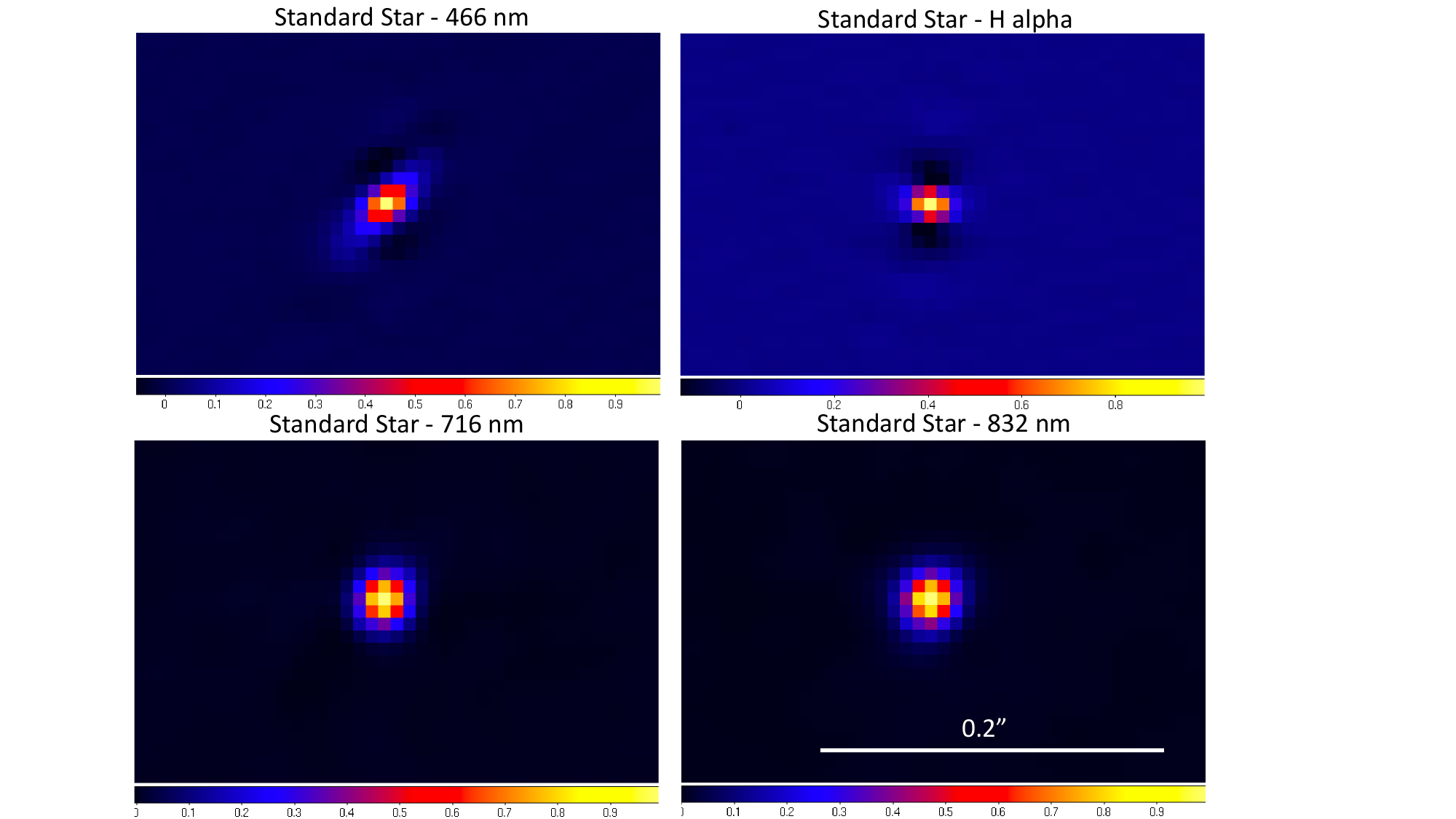}

\caption {The brightness distribution of the standard star HD 67704 as the target and HR 3179 as the PSF standard in each of the four filters of this study. The same results are obtained when the target and PSF standard are swapped. Asymmetries are evident in the H$\,\alpha$ and 466 nm images, warning of these filter images low S/N. In contrast, the brightness distributions of the standard stars in the higher-S/N images at 716 nm and 832 nm are $\sim$4 pixels wide and symmetric. North is up and East is left.
}
\label{fig:standards}
\end{figure*}

In Fig.~\ref{fig:standards} we show the brightness distribution of the standard star HD 67704 as the target and HR 3179 as the PSF standard in each of the four filters of this study. The same results are obtained when the target and PSF standard are swapped. Asymmetries (North-South in the case of the H$\,\alpha$ image and NW to SE in the 466 nm image) are evident in these images, warning of the low S/N through both these filters' images. Also shown in Fig.~\ref{fig:standards} are the standard star brightness distributions in the 716 and 832 nm images. In both cases the stars appear symmetric, with widths of $\sim$4 pixels, as expected for point sources.  We note that despite the similar total exposure times through all four filters, the 716 and 832 nm images have much higher signal-to-noise values than their 466 nm and H$\,\alpha$ counterparts.

\begin{figure*}

\hspace{0cm}
\
\includegraphics[width=11 cm]{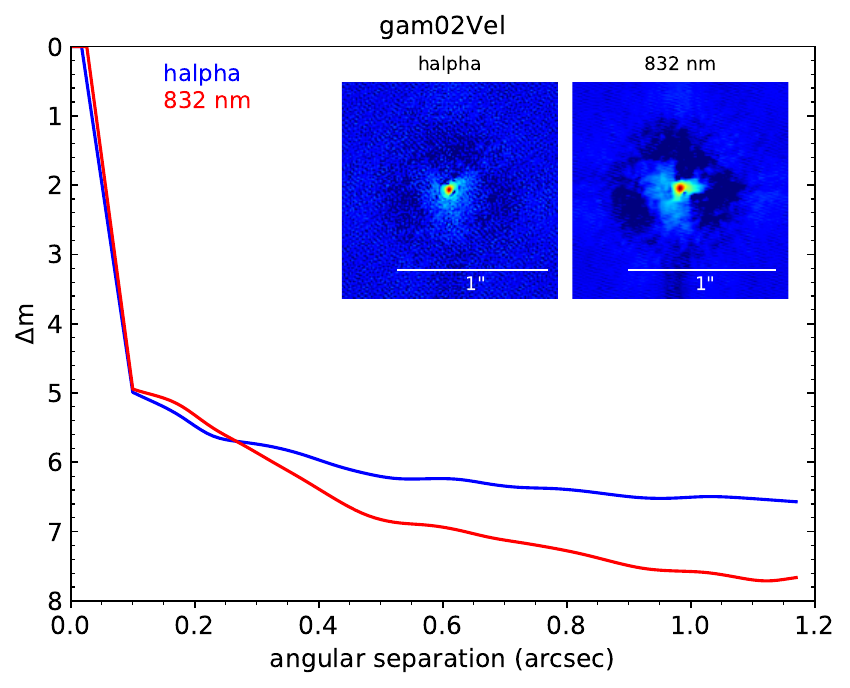}

\caption{The reconstructed speckle images and differential magnitude 5$\sigma$ detection limits $\Delta$m of the Wolf-Rayet star $\gamma^2$ Vel taken with the Zorro speckle camera through H$\,\alpha$ and 832 nm filters at Gemini-S on 2023 January 15. North is up, East is left. The 832 nm image displays an obvious $\sim$ 0.07 arcsec westward elongation.
}
\label{fig:gam2vel_ha_832}
\end{figure*}

\begin{figure*}

\includegraphics[width=11 cm]{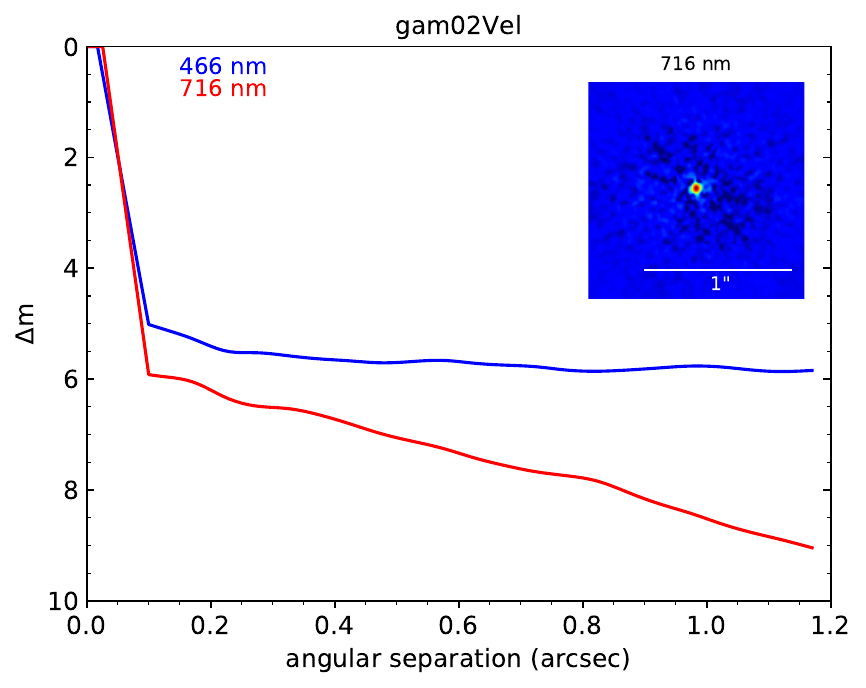}

\caption{The reconstructed speckle images and differential magnitude 5$\sigma$ detection limits $\Delta$m of the Wolf-Rayet star $\gamma^2$ Vel taken with the Zorro speckle camera through the 466 and 716 nm filters at Gemini-S on 2023 January 26. 
}
\label{fig:gam2vel_466_716}
\end{figure*}

\begin{figure*}
    \includegraphics[width=0.9\textwidth]{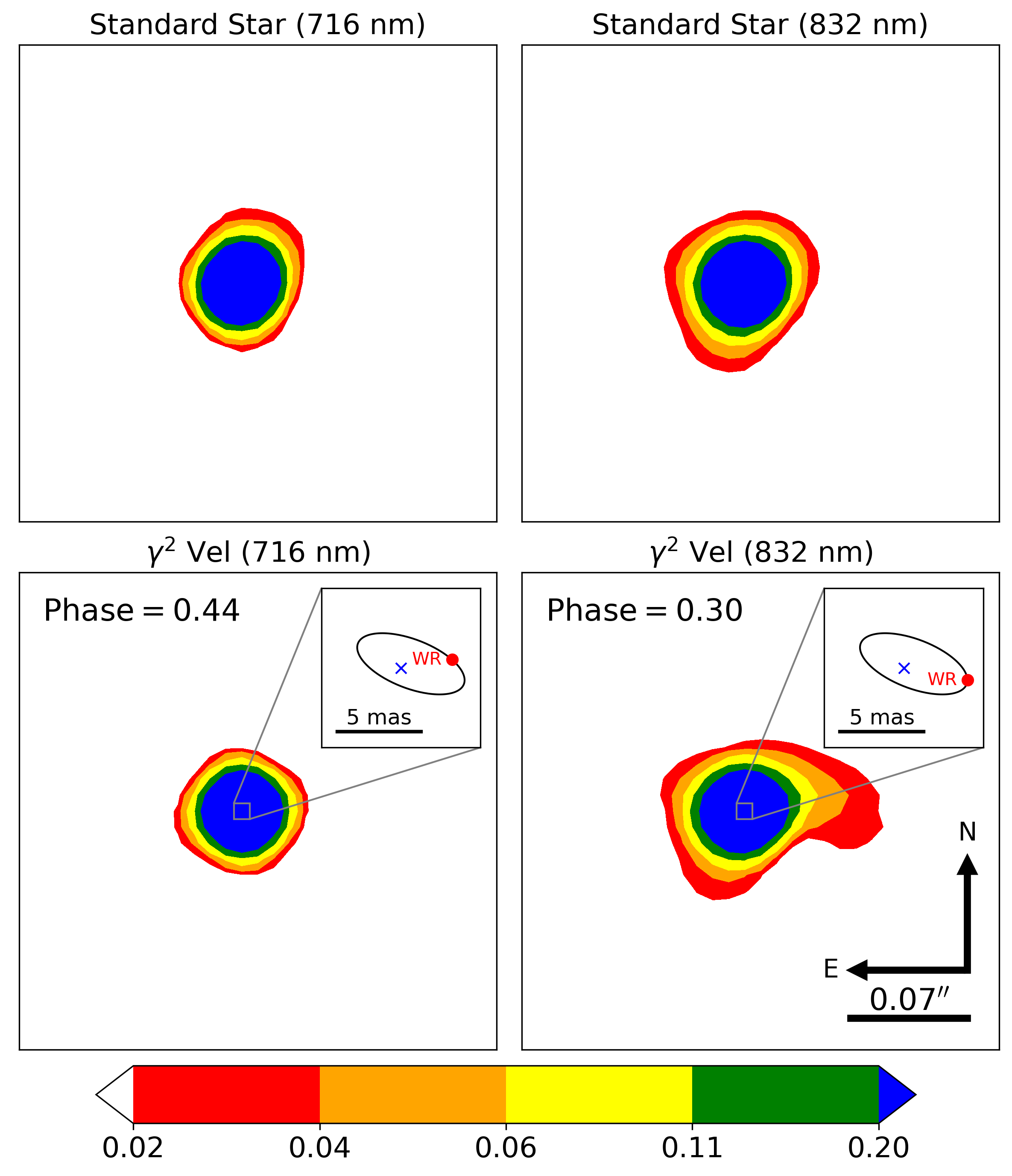}
    \caption{Contour images of the standard star (top row) and $\gamma^2$ Vel (bottom row) in the 716 nm (left column) and 832 nm (right column) filters. Brightness contours from 0.02 to 0.2 are spaced evenly on a $\log_{10}$ scale, where the brightness of the central pixel of $\gamma^2$ Vel has been normalized to 1.00.
    Regions below a brightness of 0.02 - taken to be a threshold above which features are considered significant - are shown in white, surrounding each of the red, orange, yellow green and blue contours.
    Insets in the $\gamma^2$ Vel images show the zoomed-in orbital configuration of the system at the two phases given by the parameters in \citet{North2007}. Each inset is one Zorro pixel (9.5 mas) across. The positions of the O and WR stars are marked with a blue $\times$ and red circle, respectively.
    A westward extension is seen in the 832 nm data, taken at orbital phase 0.30, but not in the 716 nm data taken at phase 0.44.}
    \label{fig:multi_contour}
\end{figure*}

The observational results on $\gamma^{2}$ Vel are shown in Figs.~\ref{fig:gam2vel_ha_832} through \ref{fig:multi_contour}.
In Figs.~\ref{fig:gam2vel_ha_832} (H$\,\alpha$ and 832 nm) and \ref{fig:gam2vel_466_716} (466 and 716 nm), the red and blue curves provide the 5$\sigma$ contrast limits between $\gamma^{2}$ Vel and any fainter companion. No such faint companions were detected. $1.5\times1.5$ arcsec reconstructed speckle images centered on $\gamma^{2}$ Vel are displayed as insets. 

The key result of this paper is that one feature is well above the noise limit and immediately evident in the 832 nm image of Figs.~\ref{fig:gam2vel_ha_832} and \ref{fig:multi_contour}: an extension stretching westward $\sim$ 0.07 arcsec from $\gamma^{2}$ Vel. The fainter ``fuzz'' to the star's east (see Fig.~\ref{fig:gam2vel_ha_832}) is near Zorro's sensitivity limit, and is possibly not real. We do not detect the 0.5 arcsec spatial extension suggested by \citet{Roche2012}, though of course our free-free continuum observations sample much denser, ``closer-in'' material than the more extended, low density matter (that allows forbidden lines) that they reported.

\section{The Asymmetry} 
\subsection{{\bf Wind illumination}}
The dominant source of light in the binary is the O star \citep{DeMarco1999}, which is $\sim$ threefold brighter at 832 nm than the WR star. The ionised wind radiates both free-free and recombination line emission determined by its state of ionisation and abundances. The densest parts of the wind - i.e. the inner regions of the WR star, and presumably the wind collision zone, will dominate in the 832 nm bandpass, but O star light may also contribute significantly. A detailed model of the wind, similar to that of \citet{Lamberts2017} but extending at least 20X farther out, is needed to directly compare with the westward extension seen in Figs.~\ref{fig:gam2vel_ha_832} and \ref{fig:multi_contour}. 
\subsection{Source of the asymmetry}
Based on previous work, there are conceivably four possible sources of asymmetry in WR (and likely all hot-star) winds: 1. Stochastic wind clumping on all scales with a likely compressible-turbulent power-law distribution \citep{Moffat1994,Lepine1999}; 2. a flattening of the wind due to rapid stellar rotation \citep{Harries2000}; 3. co-rotating interaction regions (CIRs) originating from the shocks created by semi-permanent (probably magnetically confined) bright spots on the hydrostatic stellar surface \citep{Cranmer1996,Aldoretta2016}; and/or 4. colliding-wind cone-shaped shocks in the form of corotating loose spirals in a binary system \citep{Luehrs1997,Lau2022} (because the wind speed is much greater than the rotation speed at the equator; the same applies to 3.). Possibility 2. is unlikely, since it produces only a radially smooth profile unlike what we see. Possibility 4. is not ruled out, since the extension we see occurs at the phase (0.30) where the shock cone axis is perpendicular to our line of sight \citep{Millour2007}. This leaves possibilities 1., 3., and 4. as the most likely explanations. However we do not yet have enough constraints to pin-point which of these three is responsible for the observed elongation. If 1. is correct, then one should see further random clumps in different directions at different observation times. If 3. or 4. is more likely, the observed extension depends on the phase of observation, and better data will be needed to see more of the spiral CIR or shock cone features. Further observations with better resolution, sensitivity, and at different orbital phases are needed to distinguish between the above possibilities.

\clearpage 
\clearpage

\clearpage

\clearpage

\section{Summary and Conclusions} 

We have used Gemini-S/Zorro to possibly resolve the wind of the closest Wolf-Rayet + O star binary: $\gamma^{2}$ Vel. A westward extension ($\sim$0.07 arcsec in length) of brightness is seen in the 832 nm images taken at orbital phase 0.30, with the shock cone axis perpendicular to our line of sight. The projected length of the extension is $\sim$ 24 AU, and its brightness is $\sim$a few percent that of $\gamma^{2}$ Vel. There is no hint of this extension in the 716 nm reconstructed image taken at phase 0.44, when our line of sight coincides with the shock cone axis. Higher resolution and more sensitive observations at multiple orbital phases are needed to confirm our preliminary finding, and to constrain the mechanism responsible for the observed asymmetric extension of $\gamma^{2}$ Vel.

\section*{Acknowledgments}
The data presented in this paper are based on observations
obtained at the international Gemini Observatory, a program
of NSF's NOIRLab, which is managed by the Association
of Universities for Research in Astronomy (AURA) under
a cooperative agreement with the National Science Foundation
on behalf of the Gemini Observatory partnership: the
National Science Foundation (United States), National Research
Council (Canada), Agencia Nacional de Investigación
y Desarrollo (Chile), Ministerio de Ciencia, Tecnología e Innovación (Argentina), 
Ministério da Ciência, Tecnologia, Inovações e Comunicações (Brazil), and Korea Astronomy and
Space Science Institute (Republic of Korea). This work was
enabled by observations made from the Gemini-South telescope.  
Observations in the paper made use of the High-
Resolution Imaging instrument Zorro, which was funded
by the NASA Exoplanet Exploration Program, built at the
NASA Ames Research Center by Steve B. Howell, Nic Scott,
Elliott P. Horch, and Emmett Quigley, and mounted on the
Gemini-South telescope of the international Gemini Observatory.
AFJM is grateful to NSERC (Canada) for financial
aid. We thank the Canadian Gemini Time Allocation Committee for
excellent feedback and support, and their allocation of telescope
time. The observations were obtained under Gemini
proposal GS-2022B-FT-113.

Facilities: Gemini-S

Software: \textsc{astropy} \citep{astropy:2013, astropy:2018, astropy:2022}, \textsc{numpy} \citep{harris2020array}, \textsc{poliastro} \citep{juan_luis_cano_rodriguez_2022_6817189}

\section*{Data Availability}

All relevant data, including all images and figures,
are available from the corresponding author on reasonable request.
\bibliographystyle{mnras}
\typeout{}
\bibliography{gvel}

\label{lastpage}
\end{document}